\documentclass[aps,pra,onecolumn,11pt]{revtex4}
\usepackage[utf8]{inputenc}
\usepackage{amsmath}
\usepackage{amssymb}
\usepackage{graphicx}
\usepackage[export]{adjustbox}
\usepackage{color, soul, cancel} 
\usepackage[export]{adjustbox}
\usepackage{natbib}

\begin{document}
\title{Optimal, hardware native decomposition of parameterized multi-qubit Pauli gates}
\author{P.V. Sriluckshmy}
\thanks{Corresponding author: pv.sriluckshmy@meetiqm.com}
\affiliation{IQM, Nymphenburgerstr. 86, 80636 Munich, Germany} 
\author{Vicente Pina-Canelles}
\affiliation{IQM, Nymphenburgerstr. 86, 80636 Munich, Germany}
\author{Mario Ponce}
\affiliation{IQM, Nymphenburgerstr. 86, 80636 Munich, Germany}
\author{Manuel G. Algaba}
\affiliation{IQM, Nymphenburgerstr. 86, 80636 Munich, Germany}
\author{Fedor \v{S}imkovic IV}
\affiliation{IQM, Nymphenburgerstr. 86, 80636 Munich, Germany}
\author{Martin Leib}
\affiliation{IQM, Nymphenburgerstr. 86, 80636 Munich, Germany} 
\begin{abstract}
We show how to efficiently decompose a parameterized multi-qubit Pauli (PMQP) gate  into native parameterized two-qubit Pauli (P2QP) gates minimizing both the circuit depth and the number of P2QP gates. Given a realistic quantum computational model, we argue that the technique is optimal in terms of the number of hardware native gates and the overall depth of the decomposition. Starting from PMQP gate decompositions for the path and star hardware graph, we generalize the procedure to any generic hardware graph and provide exact expressions for the depth and number of P2QP gates of the decomposition. Furthermore, we show how to efficiently combine the decomposition of multiple PMQP gates to further reduce the depth as well as the number of P2QP gates for a combinatorial optimization problem using the Lechner-Hauke-Zoller (LHZ) mapping.
\end{abstract}
\date{\today}
\maketitle
\section{Introduction}
Further accelerating the speed of scientific progress requires  computational resources beyond the capabilities of state-of-the-art classical computing. Computational power has been growing exponentially for a couple of decades according to Moore's law. However, the miniaturisation of classical computers has reached hard physical boundaries bringing Moore's Law to an end. In recent years, Quantum Computing (QC) has emerged as a promising alternative \cite{feynman_1982,shor_1999} that could provide exponentially growing compute power for application areas like quantum chemistry, optimisation and machine learning. Quantum algorithms, including speedup proofs, have been developed within all these application areas.

High-level quantum algorithms using arbitrary quantum gates need to be mapped to hardware native gates. This mapping often leads to an overhead in terms of the number of gates due to non-local and multi-qubit gates. For example, fermion-to-qubit mappings, like the Jordan-Wigner transformation \cite{jordan_1928} and others \cite{bravyi_2002,verstraete_2005,derby_2021} necessitate parameterized gates acting on more than two qubits. The encoding of complicated optimisation problems or the floating point dynamics of partial differential equations into qubits \cite{welch_2014} also typically leads to multi-qubit gates. Therefore, it is important in a majority of quantum algorithms to find optimal decompositions of multi-qubit gates into native gates.

Most QC hardware platforms do not support the direct implementation of multi-qubit gates. Building high fidelity multi-qubit interactions and inter-qubit connectivity \cite{cirac_1995, malinowski_2021, martinez_2016, lanyon_2011, gu_2017, kockum_2019, krantz_2019} has been a major hardware roadblock. Thus arises a need to search for a decomposition of the multi-qubit gates based on the native, local, single and two-qubit gates \cite{nielsen_2002,barenco_1995,molmer_1999,vartiainen_2004,mottonen_2004,mottonen_2005}. Multi-qubit gates can be decomposed into a ladder of CNOT gates and a single qubit rotation as proposed in \cite{nielsen_2002}. We argue in the present work that this is not optimal even when the CNOT gate is available as a hardware native gate. Decomposition of multi-qubit gates into two-qubit CNOT gates has also been discussed in Ref.~\cite{clinton_2021,cowtan2020}. These decompositions, however, are not symmetric, prohibiting gate cancellations between two consecutive many-body gates which we will argue can improve algorithm performance in the last part of this article.

We start this paper with  a definition of the quantum computational model. Afterwards, we propose a generalized systematic method to decompose and recursively generate parameterized multi-qubit Pauli (PMQP) gates using parameterized two-qubit gates (P2QP). We apply this method to decompose PMQP gates for some specific hardware topologies, like the path and the star graph. For the noisy intermediate scale quantum era (NISQ) \cite{preskill_2018}, the number of P2QP gates and the depth of the quantum circuit or the total run-time of the gate decomposition are key indicators of algorithmic performance. We prove that the decomposition introduced here is optimal with respect to the number of P2QP native gates as well as the overall depth. Inspired by the minimal depth proof, a procedure to decompose PMQP gate on a general hardware graph is derived. We then apply the decomposition to the four qubit Pauli gates of the parity encoded Quantum Approximate Quantum Algorithm and show further advantages of our technique with gate cancellations between different decompositions of PMQP gates.  
 
\section{Quantum Computational Model}
In order to gauge the quality of our gate decompositions we define the following quantum computational model for the remainder of the article: The computational units are geometrically separated, two-level systems, or qubits, whose states are elements in a two-dimensional complex Hilbert space. The set of possible unitary operations, or single-qubit gates, on these qubits consists of the Hadamard (H) and the S gate which can be represented in matrix form by,
\begin{align}
\text{H} &= \frac{1}{\sqrt{2}}\begin{pmatrix}
1 & 1 \\
1 & -1 
\end{pmatrix} & \text{S} &= \begin{pmatrix}
1 & 0 \\
0 & i 
\end{pmatrix}\,.
\end{align}
To create correlations and entanglement between qubits we further assume connections obeying the physical constraints of the respective hardware platform. For two qubits $a$ and $b$ that share a connection we assume that one can switch on and off a Hamiltonian of the form,
\begin{align}
    H_{\text{2q}} &= g(t) \sigma_a\sigma_b \,,
\end{align}
where $\sigma_a$($\sigma_b$) is a Pauli matrix ($\sigma \in \{x,y,z\}$), defined by the specific hardware platform, acting on qubit a(b) and $g: \mathbb{R} \rightarrow \mathbb{R}$ an arbitrary control function. With this, one can implement the following two-qubit gates,
\begin{align}
    U_{\text{2q}} = e^{i\gamma \sigma_a \sigma_b}\,,
\end{align}
for arbitrary $\gamma \in \mathbb{R}$. The single-qubit gates $\text{H}$ and $\text{S}$ can be used to rotate any Pauli matrix into any other, therefore one can implement the $U_{\text{2q}}$ two-qubit gate with arbitrary Pauli matrices $\sigma_a$ and $\sigma_b$. To fully describe the native gate set for this quantum computational model on a specific hardware platform it suffices therefore to define the hardware graph $\mathcal{G}_{\text{HW}}$ where every node corresponds to a qubit and every edge $E(\mathcal{G}_{\text{HW}})$ corresponds to a connection between the qubits. We further assume that gates that commute can be executed in parallel. While this is trivial for gates that act on non-overlapping sets of qubits, we specifically extend the notion of parallelism to gates that act on two overlapping sets of qubits. For example the two-qubit gates $e^{i \gamma_{(1,2)}z_1z_2}$ and $e^{i \gamma_{(2,3)}z_2z_3}$ can be executed in parallel, in a digital-analog fashion\cite{rodriguez_2020,yu2022}, because $H_{(1,2)}$ and $H_{(2,3)}$ can be switched on at the same time, enacting the desired combination of two-qubit gates.

The task that we want to solve is to find a decomposition of a multi-qubit gate, 
\begin{align}
    U_{\text{nq}} = e^{i\gamma P_n} = \prod\limits_{l} e^{i \beta_l\sigma_{a(l)} \sigma_{b(l)}}\,,
\end{align}
where $P_n = \sigma_1 \otimes \dots \otimes \sigma_n$ is a tensor product of $n$ Pauli matrices and the decomposition is in terms of two-qubit gates that are supported by the hardware graph, $(a(l),b(l)) \in E(\mathcal{G}_{\text{HW}})$ for all $l$. 

We further seek decompositions that are optimal with respect to a specific error model in the sense that the decomposition shows the highest possible error resilience. Current quantum computing hardware platforms are dominated by two different types of errors: finite fidelity of gate operations and the dissipative processes of the qubits themselves typically described in terms of amplitude damping and dephasing \cite{lupke_2020}. The finite gate fidelity is currently mainly due to control errors and ultimately limited by the dissipative processes of the participating qubits, therefore we aim to find a decomposition which minimizes the overall execution time of the decomposition.The parameter of the gate $\gamma$ is proportional to the time integral of the tunable interaction strength $g$. Consequentially, the parameter $\gamma$ is not necessarily related to the time it takes to implement the gate but could also be tuned by keeping the gate time fixed and changing the maximal interaction strength during gate execution. This ultimately means that even coherence time limited gates do not necessarily show a decreasing fidelity as a function of $\gamma$. In order to minimize the overall execution time we therefore have to minimize the number of parallelizable gate layers. A parallelizable gate layer consists of two-qubit gates that can be executed in parallel as discussed above. Since the execution time of a two-qubit gate is typically longer than single-qubit gates, we only count layers of two-qubit gates.

Based on the computational model developed above, a generic rule to decompose a multi-qubit gate is derived in the next section. 
\section{Recursive Construction of Gate Decompositions}
\subsection{General decomposition rule}
A general procedure to decompose a multi-qubit gate is, 
\begin{align}
U_{nq} = e^{i\gamma P_n} = e^{i \frac{\pi}{4} O_k} e^{\pm i\gamma H_l} e^{-i \frac{\pi}{4} O_k}
\end{align}
where $O_k$ and $H_l$ act non-trivially on $k<n$ and $l<n$ qubits, respectively, and they fulfill the following relations: 
\begin{align}
   \{H_l, O_k\} &= 0 & P_n &= \pm \frac{i}{2}[O_k, H_l]\,. \label{eq:constraints}
\end{align} 
If $H_l$ and $O_k$ intersect non-trivially at an odd number of qubits, i.e. they have an odd number of common qubits and the Pauli operators on odd number of these qubits don't commute, then $H_l$ and $O_k$ anticommute. If $H_l$ and $O_k$ anticommute the second equality can be fulfilled if either $iO_kH_l = P_n$ or $iH_lO_k = P_n$. This freedom of choice, as well as the specific choice of $H_l$ and $O_k$ within the requirements defined above, can be used to come up with a decomposition that has a low circuit depth considering the above defined quantum computational model.

Since both, $H_l$ and $O_k$, are also Pauli operators generating $l$-qubit and $k$-qubit gates respectively, they can be further decomposed recursively using the same decomposition rule, until all gates involved in the decomposition are native two-qubit gates.  

To simplify the description of specific decompositions in the remainder of the article we introduce a concise way to describe them. We symbolize every decomposition of an arbitrary PMQP gate generated by Pauli operator $P$ into gates generated by Pauli operators $H$ and $O$ by, $P_n \rightarrow O_{s_O}, H_{s_H}$, where $O_{s_O}$ ($H_{s_H}$) acts non-trivially on the qubits defined by the sets of nodes $s_O$ ($s_H$) and is unambiguously defined by this set. All sequences of decompositions that we describe in the following are such that all $O^{(i)}$ generate hardware native two-qubit gates and therefore all consecutive decomposition have a further decomposition of $H^{(i)}$ as a target, where $i$ denotes the sequence level of the decomposition. Consequently, a general sequence of decompositions can always be described in the following way, 
\begin{align}
    P_n \rightarrow O^{(1)}_{s_{O^{(1)}}}, H^{(1)}_{s_{H^{(1)}}} \rightarrow O^{(2)}_{s_{O^{(2)}}}, H^{(2)}_{s_{H^{(2)}}} \rightarrow \dots \rightarrow O^{(p)}_{s_{O^{(p)}}}, H^{(p)}_{s_{H^{(p)}}},
\end{align}
where we implicitly assume that $O^{(i+1)}_{s_{O^{(i+1)}}}, H^{(i+1)}_{s_{H^{(i+1)}}}$ is a decomposition of $H^{(i)}_{s_{H^{(i)}}}$, such that $s_{O^{(i+1)}} \subset s_{H^{(i)}}$ and $s_{H^{(i+1)}} \subset s_{H^{(i)}}$. The sequence of decompositions  is terminated when $H^{(p)}$, for some $p$, generates a hardware native two-qubit gate. The final decomposition of the multi-qubit gate is given as
\begin{align}
U_{nq} & = e^{i \frac{\pi}{4} O^{(1)}} \cdots  e^{i \frac{\pi}{4} O^{(p)}} e^{ \pm i\gamma H^{(p)}} e^{-i \frac{\pi}{4} O^{(p)}} \cdots  e^{-i \frac{\pi}{4} O^{(1)}}.
\end{align}
The decomposition for the path and star hardware graphs with an extension to the most general hardware graph is demonstrated in the following section. 
\subsection{Path Hardware Graph} 
\begin{figure}
\begin{center}
    \vspace{0.5cm}
\includegraphics[width=0.8\linewidth]{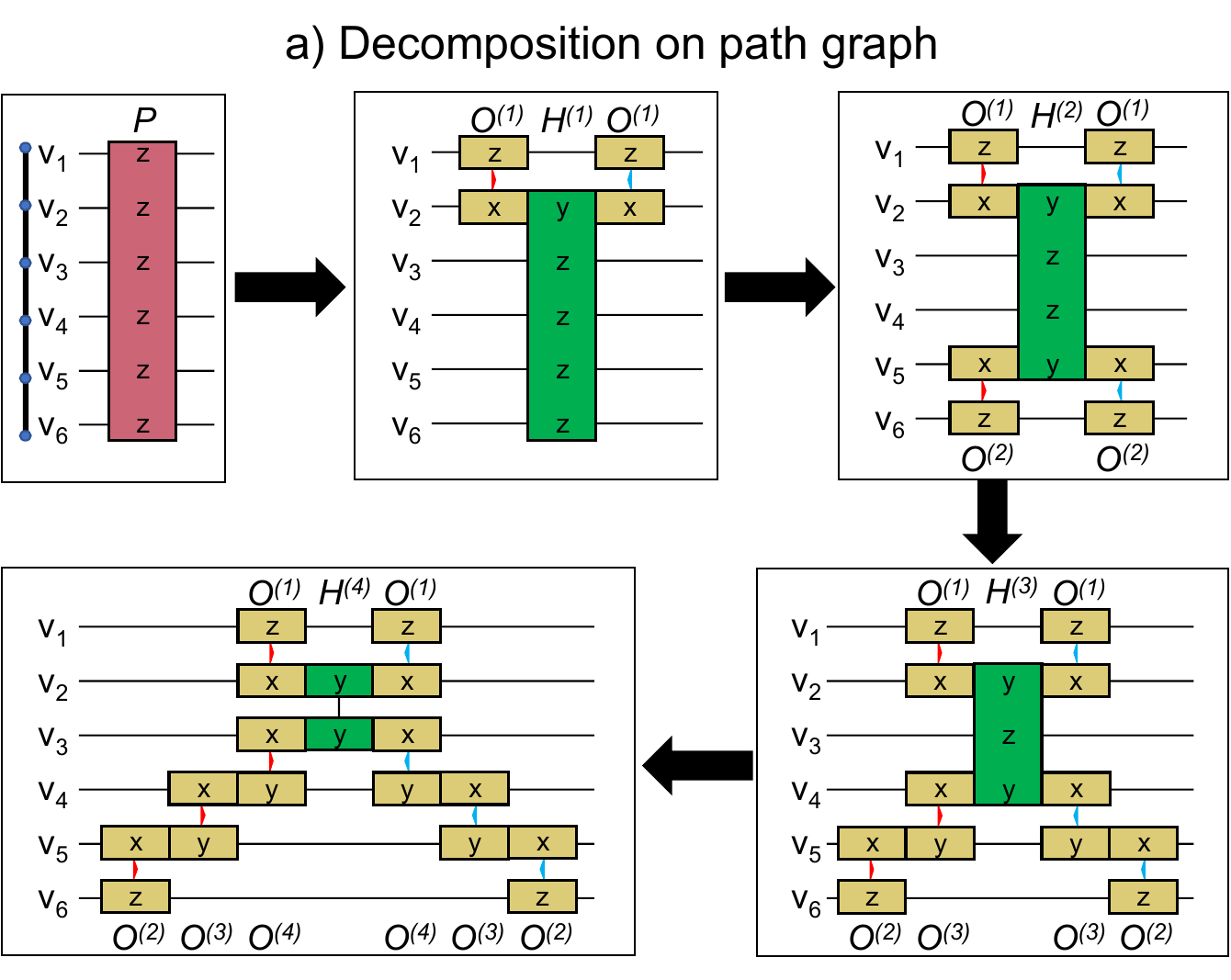}
\includegraphics[width=0.8\linewidth]{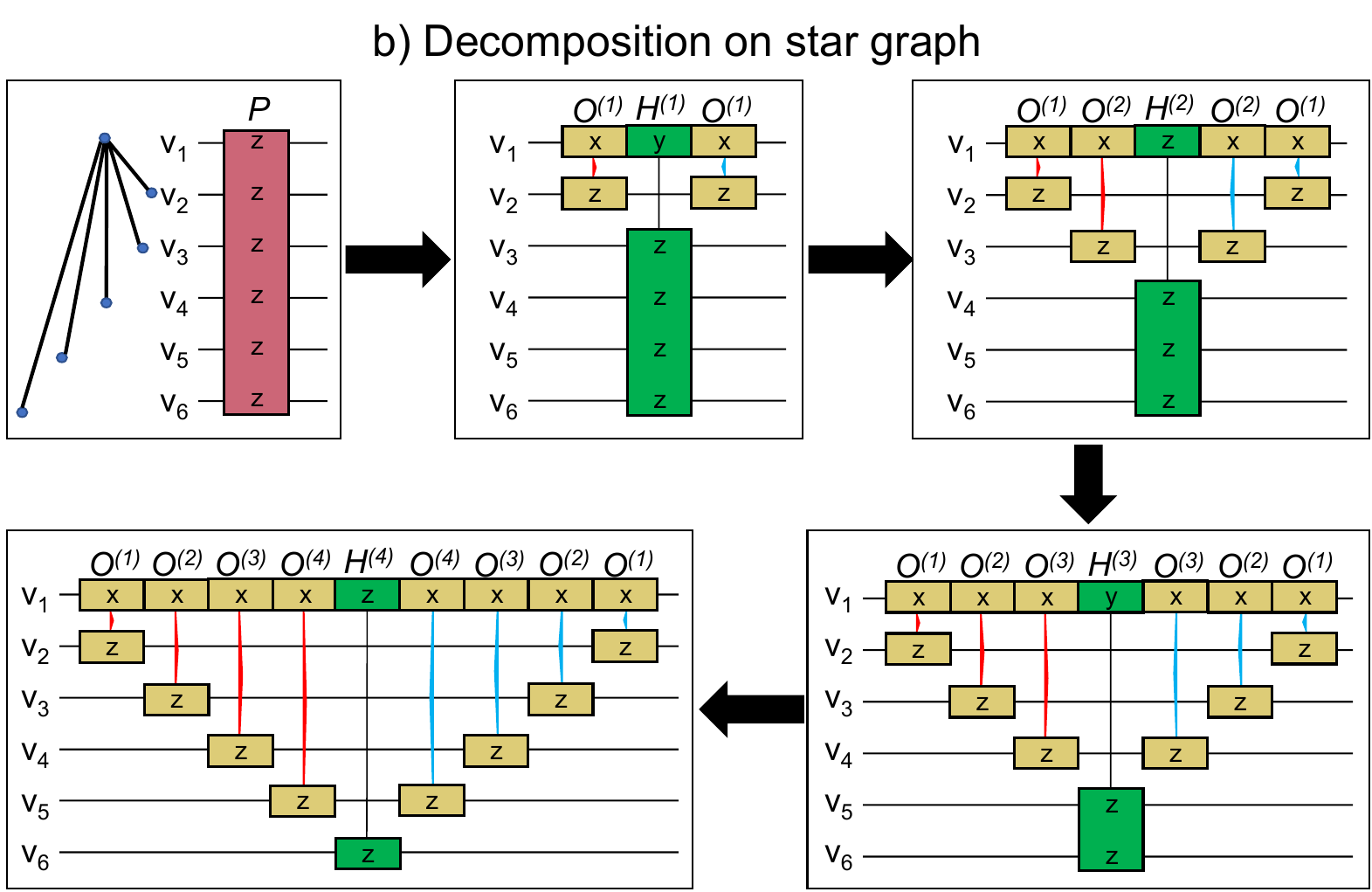}
\caption{Step-wise decomposition of a PMQP on the (a) path and (b) star graph with $6$ vertices. For the path graph, $v_m = v_2$. The yellow boxes correspond to two-qubit gates, $O^{(i)}$ and the green boxes correspond to smaller PMQP gates, $H^{(i)}$. The red and blue arrows connecting pairs of qubits represent coupling strengths of $\pm \frac{\pi}{4}$, respectively. At the end of the decomposition, only native two-qubit gates remain. We follow this color scheme through the rest of the paper.}
\label{fig:decomp_graph}
\end{center}
\end{figure}

Consider a path graph as hardware graph $\mathcal{G}_{HW}$, with $n$ vertices $v_1, v_2 \cdots, v_{n}$ and $n-1$ edges $\{(v_j, v_{j+1})|1 \leq j< n\}$, cf. Figure \ref{fig:decomp_graph}. At the first step a vertex $v_m$ from $v_1, \cdots , v_{n-1}$ is chosen. Then, the decomposition of a PMQP gate $U = e^{i\gamma P_n}$, $P_n \rightarrow O^{(1)}_{\{v_1\} \cup \{v_2\}},H^{(1)}_{\{v_2\}\cup\{v_3 \dots, v_n\}}$ is started from one of the boundaries of the path hardware graph. $v_2$ is the common connecting node which makes $O^{(1)}$ and $H^{(1)}$ anticommute. As a next step, decompose $H^{(1)}_{\{v_2, \dots, v_n\}} \rightarrow O^{(2)}_{\{v_2\}\cup\{v_3\}}, H^{(2)}_{\{v_3\}\cup\{v_4 \dots, v_n\}} $ such that the common connecting node of $O^{(2)}$ and $H^{(2)}$ is $v_3$. The process is iterated such that the index of the common node is increased by one at every step until it becomes $v_m$ as 
\begin{align}
    P_n \rightarrow O^{(1)}_{\{v_1\}\cup\{v_2\}}, H^{(1)}_{\{v_2\}\cup\{v_3 \dots, v_n\}} \rightarrow O^{(2)}_{\{v_2\}\cup\{v_3\}}, H^{(2)}_{\{v_3\}\cup\{v_4, \dots, v_n\}} \rightarrow \dots \rightarrow O^{(m-1)}_{\{v_{m-1}\}\cup\{v_{m}\}}, H^{(m-1)}_{\{v_{m}\}\cup\{v_{m+1}\dots, v_n\}},
\end{align}
occurring at step $m-1$. Afterwards, start the decomposition, beginning with node $v_{n-1}$ as the common node and decrease by one at every step to finally end at $v_{m+1}$ as
\begin{align}
    P_n & \rightarrow O^{(1)}_{\{v_1\}\cup\{v_2\}}, H^{(1)}_{\{v_2\}\cup\{ v_3\dots, v_n\}} \rightarrow O^{(2)}_{\{v_2\}\cup\{v_3\}}, H^{(2)}_{\{v_3\}\cup\{v_4 \dots, v_n\}} \rightarrow \dots 
    \rightarrow O^{(m-1)}_{\{v_{m-1}\}\cup\{v_{m}\}}, H^{(m-1)}_{\{v_{m}\}\cup\{v_{m+1} \dots, v_n\}} \nonumber\\ 
    & ~~ \rightarrow O^{(m)}_{\{v_{n-1}\}\cup\{v_{n}\}}, H^{(m)}_{\{v_{m}\}\cup\{v_{m+1}, \dots,v_{n-2}\}\cup\{ v_{n-1}\}} \rightarrow O^{(m+1)}_{\{v_{n-2}\}\cup\{v_{n-1}\}}, H^{(m+1)}_{\{v_{m}\}\cup\{v_{m+1} \dots, v_{n-3}\}\cup\{ v_{n-2}\}} \rightarrow \dots \nonumber \\ 
    & ~~~~ \rightarrow  O^{(n-2)}_{\{v_{m+1}\}\cup\{v_{m+2}\}}, H^{(n-2)}_{\{v_{m}\}\cup\{v_{m+1}\}}.
\end{align}
Every step of the decomposition adds $2$ two-qubit gates, except for the last step which adds $3$ two-qubit gates, totalling $2n-3$ two-qubit gates. The set of nodes in the hardware graph $s_{O^{(i)}}, i\leq m-1$ is distinct from the set of nodes in $s_{O^{(i)}}, i\geq m$. Therefore, every parallel layer adds $2$ two-qubit gates acting on different vertices except the central layer which contains only one two-qubit gate corresponding to $H^{(n-2)}_{\{v_{m}\}\cup\{v_{m+1}\}}$. The depth of the circuit or the number of parallel layers is $2(n-m+1)-3$ if $m < \lceil \frac{n}{2}\rceil$ and $2(m+1)-3$ otherwise. Since the quantum circuits representing the operation are not unique, choosing $v_m$ to be $v_{\lceil \frac{n}{2}\rceil}$ leads to a minimal depth of the circuit out of the equivalent decomposition strategies. Generalizing, the minimal depth can be written as $ n - mod(n+1,2)$, where $\text{mod}(x,2)$ is the modulo operation that returns the remainder of the division of $x$ by $2$. At the other extreme when $v_m$ is chosen to be either $v_1$ or $v_{n-1}$, a circuit depth of $2n-3$ is obtained. Although the depth scales linearly with the size of the Path and varies slightly for different starting vertices $v_m$, the number of two qubit gates required for all the equivalent implementations is the same, $2n-3$. 

Some explicit examples are discussed here. Firstly, consider a three qubit Pauli gate $U_{3q} = e^{i\gamma P_3}$ with $P_3 = Z_1Z_2Z_3$. From the corresponding qubit vertices $v_1, v_2, v_3$, choose $v_m = v_2$ with $P_3 \rightarrow O^{(1)}_{\{v_1\} \cup \{v_2\}},H^{(1)}_{\{v_2\}\cup\{v_3\}}$. At this level, since only two qubit gates remain, no further decomposition is required. The Pauli operators on the qubits are chosen such that the constraints in Equation \ref{eq:constraints} are satisfied. Particularly, the Pauli operators at qubit $v_2$ should commute to give $Z_2$. With the choice of $O^{(1)} = Z_1X_2$ and $ H^{(1)} = Y_2Z_3$ the decomposition 
\begin{align}
U_{3q} & = e^{i\gamma Z_1Z_2Z_3} = e^{i \frac{\pi}{4} Z_1X_2} e^{i\gamma Y_2Z_3} e^{-i \frac{\pi}{4} Z_1X_2}.
\end{align}
can be obtained. The depth of the circuit for implementing the PMQP gate is $3$.

For a four qubit PMQP gate, $U_{4q} = e^{i\gamma P_4}$ with $P_4 = Z_1Z_2Z_3Z_4$, $v_m = v_2$ is chosen. The first level of the decomposition gives $P_4 \rightarrow O^{(1)}_{\{v_1\} \cup \{v_2\}},H^{(1)}_{\{v_2\}\cup\{v_3, v_4\}} $.  Here $H^{(1)}$ is a three qubit gate and hence it is decomposed using the vertex $v_3$, using $H^{(1)}_{\{v_2\}\cup\{v_3, v_4\}} \rightarrow O^{(2)}_{\{v_3\}\cup\{v_4\}}, H^{(2)}_{\{v_2\}\cup\{v_3\}}$. Now, all the operators are two-qubit and the decomposition, maintaining the constraints,
\begin{align}
U_{4q} & = e^{i \frac{\pi}{4} Z_1X_2} e^{i \frac{\pi}{4} X_3Z_4} e^{ i\gamma Y_2Y_3} e^{-i \frac{\pi}{4} X_3Z_4} e^{-i \frac{\pi}{4} Z_1X_2}.
\end{align}
is derived.  Other equivalent decompositions to the ones presented can be obtained by choosing different vertices as $v_m$ or choosing different Pauli operators for the qubits as long as the constraints are fulfilled. Since $O^{(1)}$ and $O^{(2)}$ operate on different qubits, the depth of the circuit is still $3$. Step-wise decomposition of a six-qubit Pauli gate into two-qubit gates has been shown in Figure \ref{fig:decomp_graph}a.
\subsection{Star hardware graph}
Next, we discuss a star graph as hardware graph $\mathcal{G}_{HW}$, with $n$ vertices $v_1, v_2, \cdots ,v_{n}$ and $\{(v_1, v_{j})| 1<j\leq n\}$ as edges. This is a one-to-all connected graph. A PQMP gate $U_{nq} = e^{i\gamma P_n}$, acting on all the vertices of the star graph, can be decomposed by choosing the vertex $v_1$ as the common vertex for all recursive decompositions. The first step maps $P_n \rightarrow O^{(1)}_{\{v_1\}\cup\{v_2\}},H^{(1)}_{\{v_1\}\cup\{ v_3, \dots, v_n\}}$. We follow the steps to obtain
\begin{align}
    P_n \rightarrow O^{(1)}_{\{v_1\}\cup\{v_2\}}, H^{(1)}_{\{v_1\}\cup\{v_3, \dots, v_n\}} \rightarrow O^{(2)}_{\{v_1\}\cup\{v_3\}}, H^{(2)}_{\{v_1\}\cup\{ v_4, \dots, v_n\}} \rightarrow \dots \rightarrow O^{(m-2)}_{\{v_{1}\}\cup\{v_{n-1}\}}, H^{(m-2)}_{\{v_{1}\}\cup\{ v_n\}}.
\end{align}
The number of two-qubit gates is $2n-3$, similar to the decomposition of the Path graph. The Pauli operators on the vertex $v_1$ for all the $O^{(i)}$'s can be chosen to be the same (and hence commuting) and distinct from $H^{(n-2)}$. Therefore the depth of the circuit is in fact $3$ due to the simultaneous execution of commuting gates, as introduced in our computational model. Since every qubit is connected to the central qubit, any other order of choosing qubits, gives equivalent decompositions with the same depth cf. Figure \ref{fig:decomp_graph} b. Using the computational model defined in \cite{clinton_2022}, we obtain a logarithmic depth for a PMQP gate on an all-to-all connected graph. On the contrary, we obtain a constant depth even with an one-to-all connected graph, thanks to our computation model involving parallel gate execution. This circuit cannot be reduced further due to the structure of the technique developed here.

\subsection{Minimal Depth Proof}
We are going to derive a lower bound for the depth of a quantum circuit consisting of hardware native two-qubit gates that implement the desired multi-qubit gate. This lower bound will coincide with the depth of the decompositions found above, thereby proving their optimality as well as motivating the algorithm presented below for optimal gate decomposition of multi-qubit gates on arbitrary hardware graphs.

\begin{figure}
\begin{center}
\includegraphics[width=0.8\linewidth]{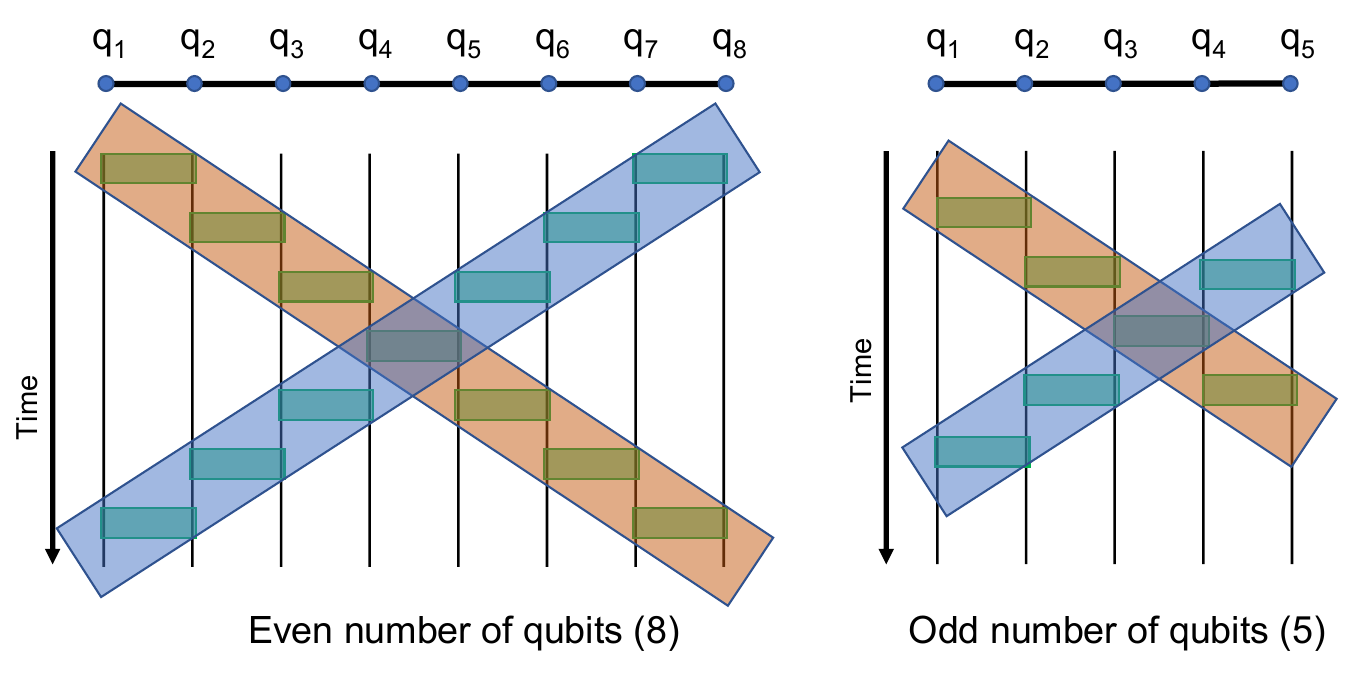}
\caption{The chain of two-qubit gates for the decomposition of a multi-qubit gate on the path graph encoded in shaded orange for $k(1) = 1$ and $k(l) = n$ and shaded blue for the reverse trajectory. The two chains overlap at one two-qubit gate. Left: $8$ vertices showing an optimal depth bound of $7$, Right: $5$ vertices showing an optimal depth bound of $5$ to entangle all the qubits.}
\label{fig:rcc}
\end{center}
\end{figure}

Without loss of generality we can assume the generator $P$ of the PMQP gate to be a tensor product of Pauli $x$ matrices on all even $n$ qubits. We obtain a lower bound of the two-qubit gate depth by proving a lower bound for the decomposition acting on an arbitrary separable state $\bigotimes_i \left|\psi_i\right\rangle $, for one specific angle $\gamma = \frac{\pi}{4}$,
\begin{align}
    \left|\Psi \right \rangle = e^{-i\frac{\pi}{4} x_1 \otimes \dots \otimes x_n } \bigotimes\limits_i\left|\psi_i\right\rangle = \frac{1}{\sqrt{2}}\left(\bigotimes\limits_i\left|\psi_i\right\rangle - i x_1 \otimes \dots \otimes x_n \bigotimes\limits_i\left|\psi_i\right\rangle \right)\,.
\end{align}
Notice that the resulting quantum state is highly correlated in the sense that all local measurements with Pauli operators that are anti-commuting with the generator of the multi-qubit gate depend on all local states $|\psi_i\rangle$ of the qubits before the gate operation,
\begin{align}
   \left\langle \Psi \right| z_j\left|\Psi\right\rangle  = \left\langle \psi_1 |x_1|\psi_1\right\rangle \dots \left\langle \psi_{j-1} |x_{j-1} | \psi_{j-1}\right\rangle \left\langle \psi_j |y_j | \psi_j\right\rangle \left\langle \psi_{j+1} |x_{j+1} | \psi_{j+1}\right\rangle \dots \left\langle \psi_n |x_n|\psi_n\right\rangle\,.
\end{align}
This implies for the two-qubit gate decomposition of the multi-qubit gate that for every ordered pair of qubits $i$ and $j$ there has to be a stair of two-qubit gates, parametrized by $k$  that connects these two qubits $U(x_{k(1)},x_{k(2)})U(x_{k(2)},x_{k(3)}) \dots U(x_{k(l-2)},x_{k(l-1)})U(x_{k(l-1)},x_{k(l)})$, such that no pair of consecutive two-qubit gates commutes, $ [U(x_{k(m-1)},x_{k(m)}),U(x_{k(m)},x_{k(m+1)})] \neq 0$, $\forall m \in \{2,\dots, l-1\}$. The depth or the number of time steps for such a decomposition is $n-1$. When the trajectory is reversed with $k(1) = j$ and $k(l) = i$, the two chains of gates overlap at $U(x_{k(\lfloor{\frac{l}{2}}\rfloor)},x_{k(\lfloor{\frac{l}{2}}\rfloor +1 )})$, cf. Figure\ref{fig:rcc}. Demanding that there is such an overlapping gate when the trajectory is reversed for an odd number of qubits, one additional gate layer is required to ensure that the decomposition is consistent. This gives a depth of $n$. For the path hardware graph this immediately leads us to the x-shaped two-qubit gate ladders that we derived above as well as the fan-shaped two-qubit patterns that we derived for the star hardware graph.

\begin{figure}
\begin{center}
\includegraphics[width=0.97\linewidth]{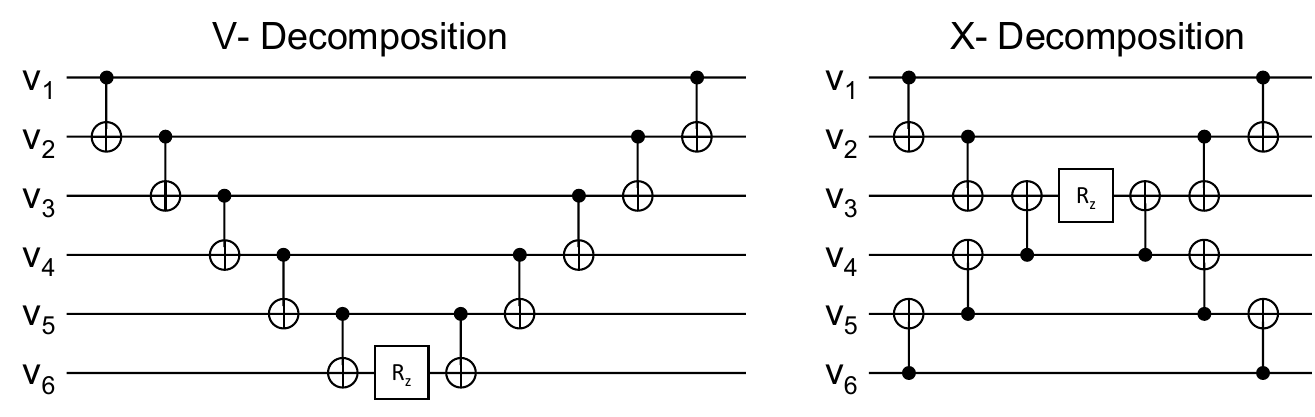}
\caption{Different decompositions of a PMQP gate on the path graph using only CNOT gates and single qubit rotations. The two-qubit gate depth of the decomposition varies from $10$ to $6$.}
\label{fig:cnot_decomp}
\end{center}
\end{figure}
Decompositions of multi-qubit gates into two-qubit gates using the CNOT gates and single-qubit rotations are in cf. Figure\ref{fig:cnot_decomp} and have also been discussed in \cite{clinton_2021}. The error model introduced in that work is dependent on the strength of the coupling and therefore the goal is to approximate a three or four qubit gate by reducing the coupling strength at the cost of increasing the number of two-qubit gates implemented. Under such a model, the decomposed circuit misses the bound by one two-qubit gate. 
\subsection{General hardware graph}
Based on the insights from the minimal depth proof above we can derive another lower bound for the depth of the decomposition on a general hardware graph. We accomplish this by identifying the longest distance between any two qubits in the hardware graph, i.e. the diameter. We define the distance between two qubits in the hardware graph, in accordance with graph theory, by the length of the shortest possible path between the two qubits. Here, a path is a sequence of vertices or qubits of the hardware graph such that consecutive vertices are neighboring  and its length is the number of edges that are traversed along the path. Between these two qubits that define the diameter of the graph there must be the aforementioned chain of two-qubit gates that consequentially lower bounds the entire depth of the decomposition of the multi-qubit gate on the given general hardware graph. We will proceed in exactly the same way as above by showing that we can find a decomposition that matches this lower bound thereby showing the optimality of the decomposition. 

Let us identify a pair of qubits with the largest possible distance in the hardware graph breaking ties arbitrarily. We define a subgraph $T$ of the hardware graph based on this found seed path graph. Add to this subgraph the shortest distance paths from every remaining qubit of the hardware graph to one of the qubits of the original set of qubits in the seed path graph. Choose the qubit in the seed path graph with the minimal distance to the current qubit, breaking ties arbitrarily. Assume, for now, that the diameter of the graph is even. Decompositions for hardware path graphs with odd diameter are a straightforward extension of the following steps. The subgraph T that we generated has now the following features: It is a rooted spanning tree, with the root being the qubit in the middle of the seed path graph. We subsume all qubits in this tree with the same distance to the root in sets that we call ``generation", where the qubits in the generation that is the furthest apart from the root are called the ``leaves". The parent for every qubit besides the root is the unique qubit it is connected to in the generation that is closer to the root qubit. The height of this rooted spanning tree, that means the longest distance between the root and any other qubit in the spanning tree $T$, is equal to half of the diameter of the hardware graph. If this would not be the case that would mean that we have identified a pair of qubits whose distance in the hardware graph is longer than the diameter of the hardware graph, which is impossible by the definition of the diameter of a graph. Lastly, all edges in the rooted spanning tree correspond to physical couplers since $T$ is a proper subgraph of the hardware graph. 

After this groundwork, we can proceed with the decomposition of the multi-qubit gate. We start with decomposing from the leaves of the spanning tree $T$: Every set of leaves together with its parent is decomposed according to the star graph decomposition, where the set of two-qubit gates generated by $O^{(i)}$ is acting on the qubits in the leaves and the respective parent qubit in the spanning tree. The remaining multi-qubit gate involves the parent qubit as well as the entire rest of the hardware graph. We iterate this procedure for every generation of qubits in the rooted spanning tree $T$ until we reach the root qubit. We finish with another star decomposition, where we choose the central gate generated by $H$ arbitrarily. If the graph has an odd diameter we would have two rooted trees connected at their roots that we identify as the spanning tree $T$. The decomposition, however, progresses in exactly the same way with the exception of the last step where the central multi-qubit gate generated by $H$ is already a two-qubit gate connecting both rooted trees that does not need to be further decomposed. 

The decomposition for every generation adds two layers of parallelizable gates to the already existing decomposition, since all involved gates can be parallelized, either because they involve disjoint pairs of qubits or are acting on the same parent qubit, however with an identical Pauli operator for the generator. We therefore managed to decompose the entire multi-qubit gate within a number of layers matching the optimal decomposition for the path hardware graph with the length given by the diameter of the general hardware graph, thereby exactly matching the lower bound identified earlier.

\begin{figure}
\begin{center}
\includegraphics[width=0.8\linewidth,valign=t]{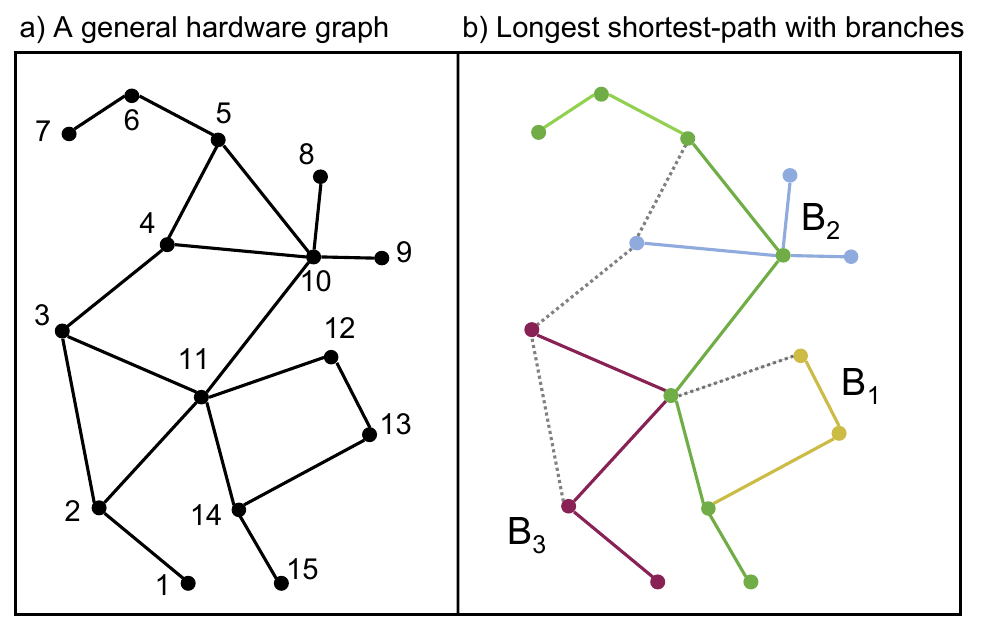}
\includegraphics[width=0.85\linewidth,valign=t]{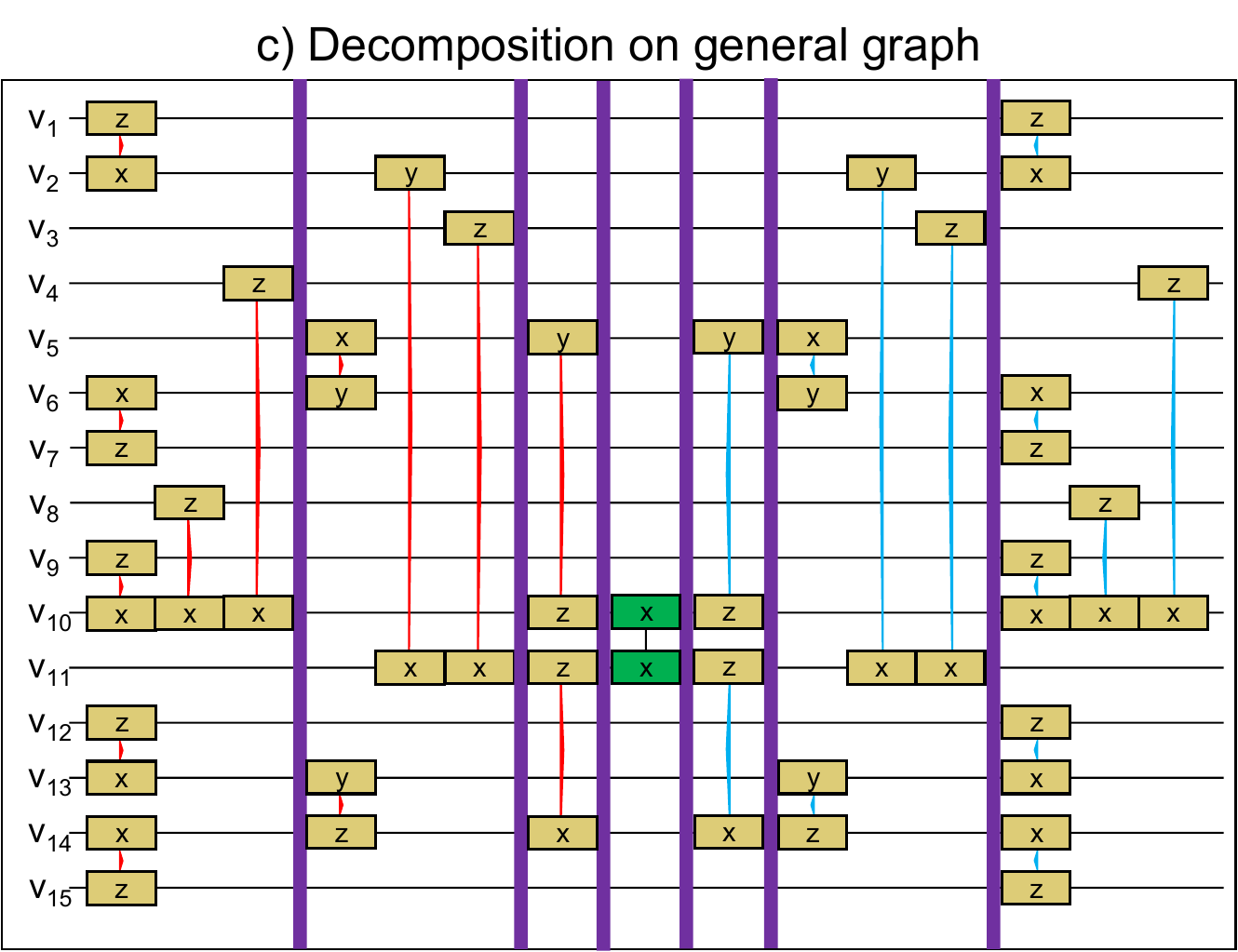}
\caption{a) A general hardware graph with $n = 15$ vertices. b) its corresponding spanning tree. The path in green is a path  with $k= 7$ vertices defining the diameter of the graph to be $6$. There are three branches $B_1$ (a local Path graph), $B_2$ (a local star graph) and $B_3$ (which is a combination of local star and path graphs).  Vertices corresponding to branches $B_1$, $B_2$ and $B_3$ are decomposed first by applying as many two-qubit gates in parallel as allowed by the computational model. Then the green vertices lying on the longest shortest-path graph are decomposed optimally giving a depth of $7$ for the decomposition. c) The final decomposition of a $U_{15q} = e^{i\gamma z_1\otimes \cdots z_{15}}$ PMQP gate on this general graph. Purple lines show the separation between the different parallel layers.}
\label{fig:generalgraph}
\end{center}
\end{figure}

In cf. Figure\ref{fig:generalgraph} we show a sample General hardware graph with $15$ vertices which requires only a depth of $7$ for its implementation.
\section{Specific example of parity encoded mapping}
\begin{figure}
\begin{center}
\includegraphics[width=0.3\linewidth,valign=t]{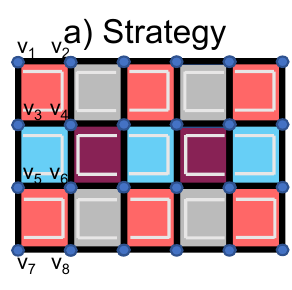}\\
\includegraphics[width=\linewidth,valign=t]{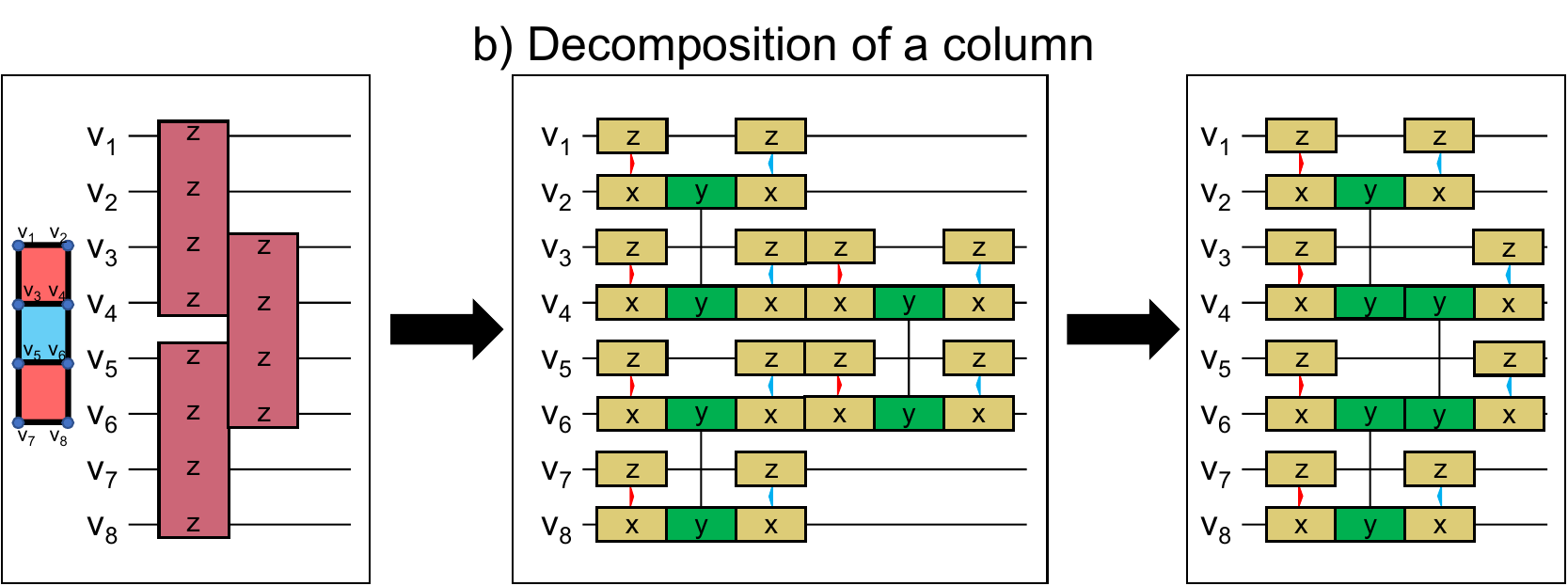}
\caption{(a) Square Grid hardware graph with the prescribed strategy for an LHZ encoded optimization problem: We choose to decompose on red squares, then blue, then gray and finally the maroon squares. The white lines within each square represents the path that we follow. (b) Decomposition of a column of four-qubit gates. Two-qubits gates acting on the same qubits with opposite signs of the coupling strengths are cancelled. Execution of all central green colored two-qubits gates in parallel results in a total depth of $3$.}
\label{fig:lhz}
\end{center}
\end{figure}

In this section, we show that consecutive decompositions of many multi-qubit gates can lead to cancellations, using a Quantum Approximate Optimisation (QAOA) circuit with a parity encoded binary optimisation problem. We do not provide the details of parity encoded QAOA, please refer to \cite{lechner_2015,lechner_2018} for the Lechner-Hauke-Zoller (LHZ) construction and \cite{ender_2021,drieb_2021,fellner_2021} for the parity architecture. For our intentions and purposes it is sufficient to know that we need to implement a gate generated by the problem Hamiltonian, 
\begin{align}
H & = \sum_i J_i \, z_i + \sum_{l}^{M} C_{l\square} \, z_{(l,n)} z_{(l,e)} z_{(l,s)} z_{(l,w)}
\end{align}
on a square grid hardware graph where $J_i$'s and $C_{l\square}$'s are constants dependent on the inital problem parameters and $n,e,s,w$ denote north, east, south, and west qubit of each plaquette ($\square$) with $M$ number of plaquettes.  The gates generated by the first term, containing local fields, can be implemented using single qubit gates in one layer. The gates generated by the second term, however, present parameterized four-qubit Pauli gates on the plaquettes of the square lattice hardware graph that require subsequent decomposition into two-qubit gates. The total run-time for the implementation of the second gate depends on the optimal decomposition of a four-qubit gate and a strategy to combine several gates that can be simultaneously executed. We follow the strategy: we choose to decompose all plaquettes with the same color in parallel, cf. Figure\ref{fig:lhz} a. We decompose the red and then blue plaquettes thereby covering all alternate columns and finally repeat the same execution to the remaining columns (gray squares and then the maroon squares). For decomposing a single four-qubit plaquette term of the form $U_{4q} = e^{i\gamma P_4} $ with $P_4 = z_1z_2z_3z_4$ we apply the protocol developed for the path graph discussed above. A simple linear Path of $v_1, v_2, v_4, v_{3}$ is chosen with $v_m = v_2$ such that the decomposition leads to 
\begin{align}
U_{4q} & = e^{i \frac{\pi}{4}z_1\,x_2}  e^{i \frac{\pi}{4}z_3\, x_4 } e^{i \gamma \, y_2\,y_4} e^{-i \frac{\pi}{4} z_3\, x_4} e^{-i \frac{\pi}{4} z_1\,x_2}.
\label{eq:four_body_eq}
\end{align} 
The same protocol can be applied to all the other plaquettes. Executing neighbouring decomposed four-qubit plaquette terms leads to a cancellation of $2$ two-qubit gates when sequentially applied to the same vertices with opposite sign of the coupling strength cf. Figure\ref{fig:lhz}. Moreover, the central two-qubit gates of all the four-qubit decompositions, namely $e^{i \gamma_{(2,4)}y_2y_4}$, $e^{i \gamma_{(4,6)}y_4y_6}$ and $e^{i \gamma_{(6,8)}y_6y_8}$, can also be executed in parallel. The two qubit circuit depth for the implementation of four-qubit plaquette terms on alternate rows is $3$ instead of $5$ which is the depth obtained using the x-shaped CNOT structure\cite{unger_2022}. Further generalization to the entire square lattice using our decomposition, can be performed in two steps of alternating rows of plaquettes giving a total constant run-time of $5$ using parallelizable commuting gates. This minimal depth of this circuit is ensured by the minimal implementation of the decomposed four-qubit gates combined with additional parallelizing strategies allowed by our computational model. This makes parity-encoded QAOA a promising problem to tackle given the currently available hardware constraints.
\section{Conclusions}
We have presented a general method to decompose PMQP into hardware implementable two-qubit gates. We demonstrated the decomposition for specific hardware graphs: the Path graph and the star graph. We show that the lower bound for the depth of the decomposition is set by the correlation of qubits for a multi-qubit gate. Further, we show that our decomposition can achieve this bound, scales linearly for the path graph and is constant for the star graph. Therefore, the less connected the graph is, the more enhanced the depth of the circuit becomes. Motivated by the minimal depth proof, we provide a strategy to optimally decompose a multi-qubit gate on any general hardware graph. For a specific quantum circuit for combinatorial optimization using the LHZ mapping, we show that the lowest depth that can be achieved is $6$, independent of the size of the system.  The technique also presents an efficient way to enable the decomposition of long-range multi-qubit interactions. For Hamiltonian systems with many multi-qubit terms, sub-optimal decompositions of some of the multi-qubit gates could be more beneficial and further facilitate gate cancellation strategies. While we present only a few use cases, the decomposition is universal and can be used to provide low-depth circuits for a wide range of near-term quantum applications. In a recent publication \cite{algaba_2023} some of the authors use the decomposition technique for fermionic systems and develop additional strategies of gates along with an optimal fermion-to-qubit mapping to reduce the depth of the circuit further. Reducing the depth reduces errors and therefore helps in developing better noise mitigation strategies.These are crucial, but not limited to the NISQ era with minimal computational effort and the hardware facilities currently available.
\section*{Acknowledgments} The authors would like to thank Inés de Vega, Hermanni Heimonen, Bruno G. Taketani and Mikko M\"{o}tt\"{o}nen for useful discussions. This project is supported by the Federal Ministry for Economic Affairs and Climate Action on the basis of a decision by the German Bundestag through the project Quantum-enabling Services and Tools for Industrial Applications (QuaST).


\end{document}